\documentclass[10pt,twocolumn,letterpaper]{article}
\usepackage[T1]{fontenc}
\usepackage{preprint}
\usepackage{amsthm}
\usepackage{algorithm}
\usepackage{algpseudocode}
\usepackage{array}
\usepackage{tabularx}
\usepackage{tikz}
\usetikzlibrary{arrows.meta,positioning}

\theoremstyle{definition}

\newcommand{\Compress}{\mathsf{Compress}}
\newcommand{\Decompress}{\mathsf{Decompress}}

\newcommand{\frameworkfigref}[2]{%
  \ifcsname r@#1\endcsname
    Fig.~\ref{#1}%
  \else
    \textcolor{black}{Fig.~[#2]}%
  \fi}
\newcolumntype{Y}{>{\raggedright\arraybackslash}X}
\definecolor{linkblue}{rgb}{0.21,0.49,0.74}
\usepackage[breaklinks,colorlinks,allcolors=linkblue]{hyperref}
\hypersetup{
  pdftitle={Lossless Reversible Watermarking for AIGC with Applications in Authentication},
  pdfauthor={Yupeng Qiu, Han Fang, Ee-Chien Chang},
  pdfsubject={Lossless reversible watermarking and authentication for AI-generated images}
}

\newcommand{\auditrev}[1]{#1}

\makeatletter
\providecommand{\theHALG@line}{}
\renewcommand{\theHALG@line}{\thealgorithm.\arabic{ALG@line}}
\makeatother
\title{Lossless Reversible Watermarking for AIGC with Applications in Authentication}
\author{
Yupeng Qiu$^{1}$\qquad Han Fang$^{2}$\qquad Ee-Chien Chang$^{1}$\\[0.5em]
{\normalsize $^{1}$School of Computing, National University of Singapore, Singapore}\\
{\normalsize $^{2}$School of Cyber Science and Technology,}\\
{\normalsize University of Science and Technology of China, Hefei, Anhui, China}
}
\date{}
\begin{document}
\maketitle
\begin{abstract}
AIGC watermarking is commonly used for tracing, but it can also support applications such as authentication and tampering detection. We study lossless reversible watermarking, which enables exact recovery of the original watermarked image after permissible distortions. By exploiting the relatively low entropy of the information required for image generation, together with robust latent-space watermarking, we design a framework that preserves sufficient information for lossless recovery. We then apply the framework to AIGC authentication, which generally faces two key challenges. First, the authentication mechanism must resist adaptive adversaries. Second, the content relevant to authentication, such as visual designs or expressed messages, may depend on application requirements that are not known at generation time. To solve these difficulties, we incorporate \auditrev{cryptographic message authentication} to provide unforgeability against adaptive attacks. Lossless recovery provides the original image as reference evidence, after which a suitable analyzer can be selected to make the authentication decision. Experiments on Stable Diffusion demonstrate exact recovery under a wide range of distortions.
\end{abstract}

\section{Introduction}
\label{sec:introduction}

{\color{black}
Diffusion models have made high-fidelity image generation widely accessible across diverse applications~\cite{podell2024sdxl,rombach2022ldm,saharia2022imagen}. As generated images are increasingly shared, edited, and reused, verifying their authenticity requires distinguishing permissible transformations from modifications that invalidate the original content claims~\cite{arabi2025seal,zhang2025omniguard,zhang2026sembind}. For example, replacing a person's face may invalidate an identity claim even if the watermark remains detectable.

Watermark-based image authentication has been extensively studied in traditional settings~\cite{fridrich1999selfembedding,lin2001sari}. Recent approaches have extended this paradigm to AI-generated images, with some methods binding watermarks to predefined semantic representations of the generated content~\cite{arabi2025seal,zhang2026sembind}. However, two fundamental challenges remain. First, robust watermark recovery does not necessarily guarantee resistance to adaptive forgery attacks, which may alter critical image content while preserving valid watermark information~\auditrev{\cite{gao2026csi,jain2025forging,muller2025forgery}}. Second, authentication requirements are application-dependent and may not be known at generation time. For instance, in an image containing a person and a logo, modifying the logo may be acceptable for verifying the person's identity but unacceptable for verifying the logo's integrity. A fixed semantic representation may therefore discard information essential to subsequent verification tasks.

These challenges raise two questions. First, how can watermark-based authentication resist adaptive forgery attacks without equating successful watermark recovery with content authenticity? Second, how can a watermark support authentication requirements that are not specified at generation time?
\par}

{\color{black}
To address these challenges, we propose a \textbf{lossless reversible watermarking {\color{black}(LRW)} framework} that enables exact recovery of the original watermarked image for downstream authentication. Rather than committing to a predefined representation of image content, our framework preserves the information needed to recover the original image. The recovered original image then serves as a reference for application-specific verification, allowing downstream applications to determine which content properties to authenticate and which transformations to accept.

\textbf{For robustness}, our framework builds on latent-space diffusion watermarking~\cite{chang2026maxmark,gunn2025undetectable,hu2026spherical,yang2025t2smark,yang2024gaussian}. These methods embed information into the initial noise and recover it through diffusion inversion and decoding, enabling high-capacity watermarking while preserving generation quality. Prior work has demonstrated robustness against common image distortions~\cite{yang2025t2smark,yang2024gaussian}, diffusion-based purification~\auditrev{\cite{gunn2025undetectable,hu2026spherical}}, and instruction-guided editing~\auditrev{\cite{hu2026spherical}}. We use this robustness to extract the generation information needed for exact image recovery, rather than relying on watermark detection alone to determine content authenticity.

{\color{black}
\textbf{For reversibility}, we preserve the inputs needed to regenerate the original watermarked image. A diffusion model generates an image from a prompt and initial Gaussian noise sampled using a random seed. We losslessly compress the prompt~\cite{deletang2024compression} and combine it with the seed, decoding metadata, and authentication information to form a watermark message embedded in the initial noise. During verification, the recovered message enables the verifier to restore the original prompt and reconstruct the exact watermarked initial noise. These inputs reproduce the original watermarked image under the same generation settings.\par}

\textbf{For authenticity}, we protect the watermark message using a cryptographic message authentication code (MAC)~\cite{bellare1996keying,bellare2004verification}. \textcolor{black}{Given a received image, the verifier first extracts the message and validates its MAC. If the MAC is valid, it recovers the original image and proceeds to content-level verification. An application-specific distance metric or analyzer then compares the recovered original with the received image to assess the relevant authenticity claims.} The MAC protects the recovery information against forgery, while robust watermarking enables its recovery under permissible distortions. This separation of image recovery and content-level verification supports application-specific authentication without predetermining which image properties to protect at generation time.

Our contributions are summarized as follows:
\begin{itemize}
    \item We propose a lossless reversible watermarking framework that separates original image recovery from downstream content verification. By recovering the whole original image rather than a limited set of predefined semantic features, our framework provides comprehensive reference evidence for authentication requirements specified after generation.
    \item We design a compact watermark message and a \textcolor{black}{reproducible} embedding procedure that enable exact image recovery without maintaining a per-image archive. Cryptographic authentication further protects the recovery information against adaptive adversarial attacks.
    \item We conduct extensive experiments across multiple watermarking methods, datasets, and image transformations, demonstrating the effectiveness of our framework in achieving lossless recovery of the original watermarked image and highlighting its potential for broader downstream authentication applications.
\end{itemize}
\par}

\section{Related Work}
\label{sec:related}

\subsection{Latent-space diffusion watermarking}
Text-to-image latent diffusion models generate images from initial Gaussian noise sampled by a seeded pseudorandom generator. A text encoder converts the input prompt into conditioning features that guide iterative denoising, transforming the initial noise into a clean latent representation, which is then decoded into an image~\cite{rombach2022ldm}. This process can be approximately reversed through diffusion inversion, which estimates the initial noise from a generated image. Latent-space watermarking exploits this property by embedding information into the initial noise and recovering it through diffusion inversion and watermark decoding.

Tree-Ring~\cite{wen2023treering} introduces structured Fourier-domain patterns for watermark detection. Subsequent methods extend this approach to multi-bit message embedding. Gaussian Shading~\cite{yang2024gaussian} encodes messages through Gaussian interval sampling, while PRC~\cite{gunn2025undetectable} employs pseudorandom error-correcting codes to improve robustness and security. More recent methods explore tail-truncated sampling with session-key randomization~\cite{yang2025t2smark}, binary message mappings through spherical rotation and radial scaling~\cite{hu2026spherical}, and sign-based embedding with learned invertible transformations~\cite{chang2026maxmark}. \textcolor{black}{These methods provide robust message embedding and recovery, but reliable content authentication requires an appropriate framework for using the recovered information. Our framework uses this capability to recover the original image as reference evidence for downstream authentication. Limited capacity further motivates lossless text compression.}

\begin{figure*}[t]
\centering
\includegraphics[width=0.90\textwidth]{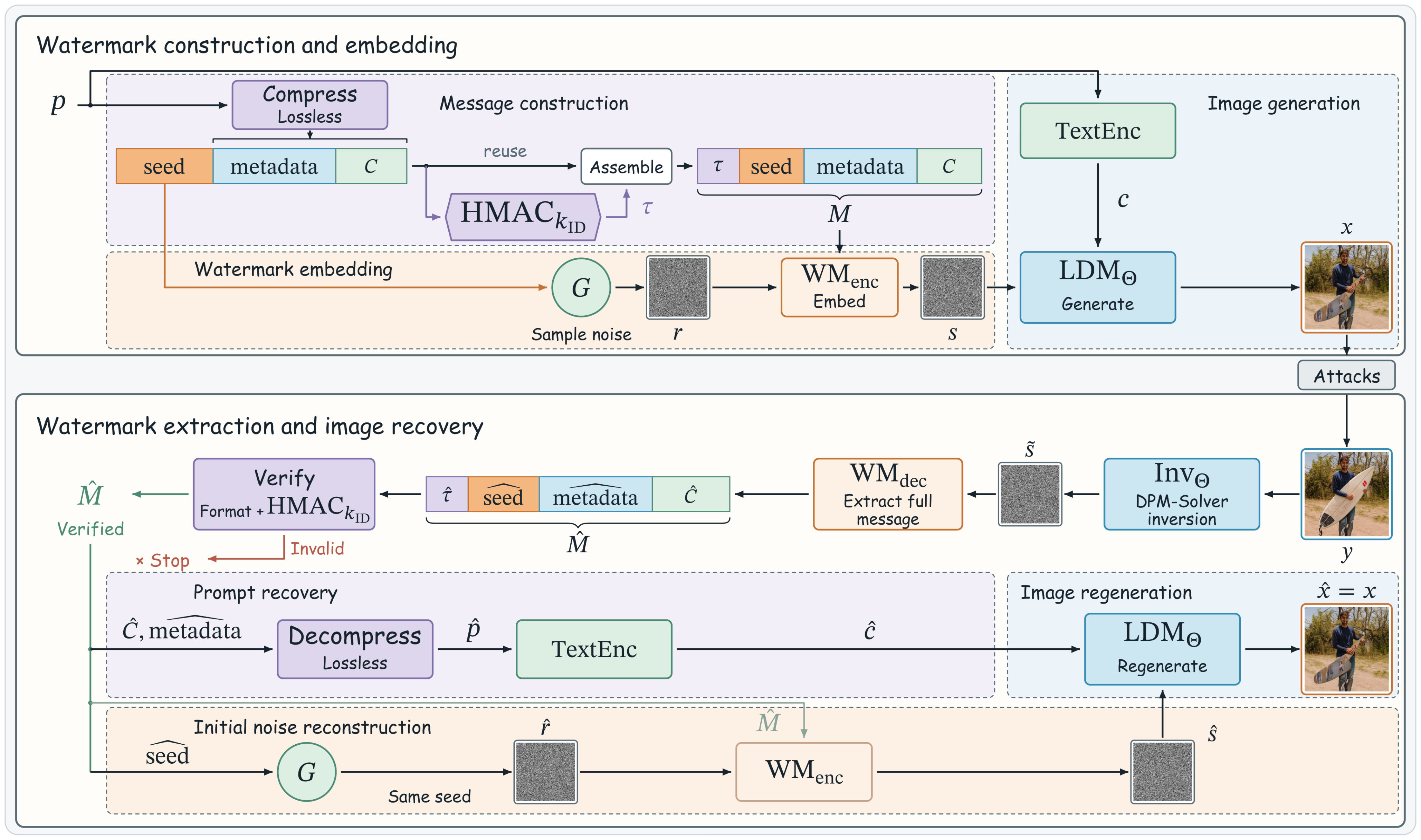}
\caption{\textbf{Lossless Reversible Watermarking (LRW).} The sender embeds the compressed prompt, seed, decoding metadata, and HMAC tag as a watermark message. The receiver verifies the extracted message, reconstructs the prompt and watermarked initial noise, and recovers the original image.}
\label{fig:framework}
\end{figure*}

\subsection{Lossless text compression}
Images typically contain high-dimensional information and require substantial storage even after lossless compression. In contrast, natural language exhibits strong statistical regularities that enable efficient compression. Text-to-image diffusion models connect these two \textcolor{black}{modalities} by generating high-dimensional images from textual prompts and random seeds. Under fixed generation settings, this allows an image to be reconstructed from its generation inputs rather than its pixel data, providing an opportunity for lossless image recovery with substantially less information.

Lossless text compression exploits statistical dependencies in language to reduce storage while preserving the original text exactly. Traditional methods combine probabilistic prediction with entropy coding~\cite{cleary1984compression,witten1987arithmetic}, while neural approaches improve compression through learned language models. Representative methods include DeepZip~\cite{goyal2019deepzip}, which uses recurrent networks, and \textcolor{black}{Language Modeling Is Compression}~\cite{deletang2024compression}, which demonstrates the compression capabilities of pretrained language models. Subsequent work further improves compression through model adaptation and efficient coding strategies, including FineZip~\cite{mittu2024finezip}, LMCompress~\cite{li2025lmcompress}, and test-time probability adjustment~\cite{zhang2025steering}. Building on these advances, our framework employs lossless text compression to reduce the watermark payload required to preserve the original prompt for exact image recovery.

\subsection{Adaptive attacks and message authentication}
Robust watermarks may remain detectable even after adversarial modifications to image content, enabling forgery attacks. A representative example is reprompting, which reuses estimated watermarked noise with a different prompt to generate new content~\cite{muller2025forgery}. To mitigate such attacks, semantic-aware methods bind watermarks to image content. SEAL~\cite{arabi2025seal} derives watermark patterns from caption-based features, while SemBind~\cite{zhang2026sembind} introduces image-dependent semantic masks. \auditrev{However, these methods rely on learned semantic representations. Targeted attacks can manipulate image captions~\cite{chen2018showandfool}, while semantic injection can bypass SEAL's verification~\cite{gao2026csi}.}

Cryptographic techniques have also been explored to protect watermark information. MetaSeal~\cite{zhou2026metaseal} embeds digitally signed image captions, while a theoretical construction for robust publicly detectable watermarking~\cite{fairoze2025difficulty} combines cryptographic signatures with image embeddings. In shared-secret settings, message authentication codes (MACs), such as HMAC, provide message integrity and authentication with security against adaptive message forgery~\cite{bellare1996keying,bellare2004verification}. Our framework uses HMAC to authenticate the generation information required for lossless reversible watermarking, enabling secure recovery of the original image as reference evidence for downstream applications.

\section{\textcolor{black}{Method}}
\label{sec:method}

{\color{black}Our method separates lossless image recovery from downstream content authentication. Lossless Reversible Watermarking (LRW) exploits the low entropy of generation information to enable exact recovery of the entire original image from a compact watermark. LRW-based authentication (LRW-Auth) then uses the recovered original as reference evidence to enable verification under the requirements of different applications.}

\begin{figure*}[t]
\centering
\includegraphics[width=0.98\textwidth]{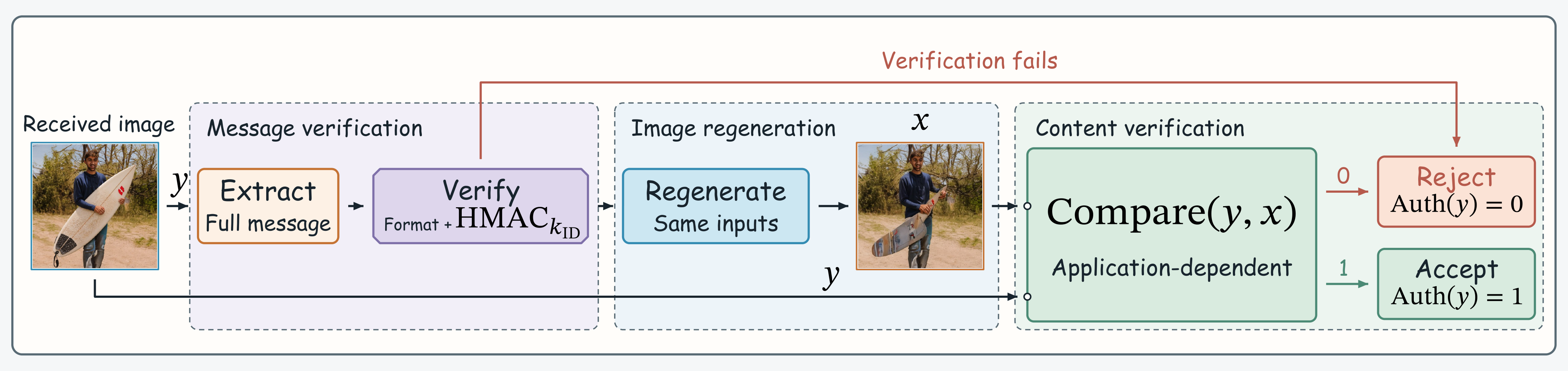}
\caption{\textcolor{black}{\textbf{LRW-based authentication (LRW-Auth).} LRW extracts and verifies the watermark message to recover the original image $x$. Failed recovery leads to rejection. Otherwise, $\mathrm{Compare}(y,x)$ evaluates the received image $y$ against $x$ according to the application's authentication requirements, returning $1$ for acceptance and $0$ for rejection.}}
\label{fig:lrw-auth-framework}
\end{figure*}

\subsection{\textcolor{black}{Lossless Reversible Watermarking (LRW)}}
\label{sec:lrw}
\label{sec:formulation}

{\color{black}LRW embeds the losslessly compressed prompt, seed, decoding metadata, and authentication information into the initial noise as a watermark message (\frameworkfigref{fig:framework}{LRW framework}). After extracting and verifying this message, the verifier recovers the prompt and reconstructs the watermarked initial noise to regenerate the original image.}

\subsubsection{Watermark Construction and Embedding}
\label{sec:wm-generation}
{\color{black}
{\color{black}Watermark construction aims to preserve the information required for exact image recovery within a compact, authenticated message. Seed-based noise sampling and lossless prompt compression provide a compact representation of the generation information, while HMAC protects its integrity. An existing latent-space watermarking method embeds this message during image generation. The following paragraphs describe these components and their integration.}

{\color{black}
\paragraph{Initial noise sampling.}
An LDM starts from Gaussian noise. We denote the seeded Gaussian sampler by $G$~\cite{lee2026authorship}:
\begin{equation}
    r = G(\mathrm{seed}).
\label{eq:initial-noise-sampling}
\end{equation}
Here, $r$ is Gaussian noise sampled from $\mathcal{N}(0,I)$. The same seed produces the same noise, allowing it to be regenerated during image recovery.
\par}

{\color{black}
\paragraph{Lossless prompt compression.}
We use a large language model (LLM) and arithmetic coding to losslessly compress the prompt $p$~\cite{deletang2024compression,witten1987arithmetic}. The LLM supplies token probabilities to the arithmetic coder, which produces a {\color{black}variable-length} binary code $C$. During encoding, the sender records the prompt's token count and the bit length of $C$ as $\mathrm{metadata}$ to support subsequent decoding. We write
{\setlength{\belowdisplayskip}{0pt}\setlength{\belowdisplayshortskip}{0pt}
\begin{equation}
\textcolor{black}{\Compress(p) = (C,\mathrm{metadata}).}
\label{eq:prompt-compression}
\end{equation}
\par}
\par}

{\color{black}
\paragraph{Message format and authentication.}
Each sender shares a user-specific secret authentication key $k_{\mathrm{ID}}$ with the verifier. We authenticate the metadata, seed, and compressed prompt $C$ using HMAC~\cite{bellare1996keying,bellare2004verification} and include the resulting tag $\tau$ in the watermark message:
\begin{equation}
\begin{aligned}
    \tau &= \mathrm{HMAC}_{k_{\mathrm{ID}}}
    (\mathrm{metadata}\mathbin\Vert\mathrm{seed}\mathbin\Vert C),\\
    M &= \underbrace{\tau\mathbin\Vert\mathrm{metadata}\mathbin\Vert\mathrm{seed}}_{\text{fixed-length part}}
    \mathbin\Vert\underbrace{C}_{\makebox[0pt]{\scriptsize variable-length part}}.
\end{aligned}
\label{eq:generation-message}
\end{equation}
Here, $\Vert$ denotes concatenation. \textcolor{black}{$M$ is the complete watermark message to be embedded into the initial noise.}
\par}

{\color{black}
\paragraph{\textcolor{black}{Watermark embedding and image generation.}}
Our framework can use any latent-space watermarking method that supports embedding and recovery of the message $M$~\cite{yang2024gaussian,gunn2025undetectable,yang2025t2smark}. \textcolor{black}{The text encoder converts the prompt $p$ into conditioning $c=\mathrm{TextEnc}(p)$.} {\color{black}We denote its watermark embedding operation by $\mathrm{WM}_{\mathrm{enc}}$ and generate the watermarked image as follows:}
{\color{black}
\begin{equation}
\begin{aligned}
    s &= \mathrm{WM}_{\mathrm{enc}}(M,r),\\
    x &= \mathrm{LDM}_{\Theta}(s,c).
\end{aligned}
\label{eq:wm-generation}
\end{equation}
\par}
\textcolor{black}{Here, $\Theta$ denotes the LDM parameters.}
{\color{black}The encoder embeds the message $M$ into the Gaussian noise $r$ to obtain the watermarked initial noise $s$. The LDM then generates the original watermarked image $x$ from $s$ and the prompt conditioning $c$.}
\par}
\par}

\subsubsection{Watermark Extraction and Image Recovery}
\label{sec:image-recovery}
{\color{black}Image recovery aims to reconstruct the original watermarked image $x$ from a possibly distorted observation $y$. Watermark extraction recovers the generation information, while HMAC verifies its integrity before reconstruction. Lossless prompt decompression and reconstruction of the watermarked initial noise provide the inputs needed to regenerate $x$ under the sender's generation settings. The following paragraphs describe these components and their integration.}

\paragraph{Complete message extraction.}
The verifier encodes $y$ \textcolor{black}{back} into latent space and applies \textcolor{black}{DPM-Solver inversion} with an empty prompt~\textcolor{black}{\cite{hong2024exact}}. Denoting the estimated initial noise by $\tilde s$, we write the extraction process as
\begin{equation}
    \tilde s=\mathrm{Inv}_{\Theta}(y),\qquad
    \hat M=\mathrm{WM}_{\mathrm{dec}}(\tilde s).
\label{eq:wm-extraction}
\end{equation}
Using the shared watermark configuration, the decoder first extracts the fixed-length part, obtaining $\hat\tau$, $\widehat{\mathrm{metadata}}$, and $\widehat{\mathrm{seed}}$. {\color{black}It then reads the code length from $\widehat{\mathrm{metadata}}$ and extracts the specified number of bits as the variable-length code $\hat C$.} Together, these fields form the complete message $\hat M$ in the format of Eq.~\ref{eq:generation-message}.

\paragraph{Message verification.}
After the complete message has been extracted, the verifier checks its format and uses the sender's authentication key $k_{\mathrm{ID}}$ to check whether the following equality holds:
\begin{equation}
    \hat\tau=
    \mathrm{HMAC}_{k_{\mathrm{ID}}}
    (\widehat{\mathrm{metadata}}\mathbin\Vert\widehat{\mathrm{seed}}\mathbin\Vert\hat C).
\label{eq:message-verification}
\end{equation}
If extraction, parsing, or tag verification fails, the procedure reports failure and stops. Otherwise, the authenticated message is used for image regeneration.

\paragraph{Image regeneration.}
\textcolor{black}{Using the same tokenizer, LLM, and coding rules as the sender, the verifier decompresses the recovered code to obtain the prompt:}
\begin{equation}
    \textcolor{black}{\Decompress(\hat C,\widehat{\mathrm{metadata}})=\hat p.}
\label{eq:recover-inputs}
\end{equation}
The verifier then regenerates the Gaussian noise $\hat r=G(\widehat{\mathrm{seed}})$ and obtains the prompt conditioning $\hat c=\mathrm{TextEnc}(\hat p)$. Repeating the sender's watermark embedding and image generation gives
\begin{equation}
\begin{aligned}
    \hat s &= \mathrm{WM}_{\mathrm{enc}}(\hat M,\hat r),\\
    \hat x &= \mathrm{LDM}_{\Theta}(\hat s,\hat c).
\end{aligned}
\label{eq:image-reconstruction}
\end{equation}
{\color{black}Diffusion inversion is used only to extract the message. If the original message is correctly recovered, the verifier recovers the same prompt and reconstructs the same watermarked initial noise. Repeating image generation under the same model, sampling settings, and numerical execution then recovers the original image exactly, $\hat x=x$. We denote this recovery procedure by $\mathrm{Recover}_{k_{\mathrm{ID}},\Theta}(y)$.}

{\color{black}
\subsection{\textcolor{black}{LRW-Based Authentication}}
\label{sec:application-authentication}

\textcolor{black}{LRW-based authentication (LRW-Auth) aims to support diverse authentication requirements without fixing them at generation time. LRW serves as the image-recovery module, providing the original image as reference evidence. An application-defined comparison determines which content to authenticate and what transformations are acceptable. The following paragraph describes how these components determine the authentication result.}

\paragraph{Authentication framework.}
\label{sec:authentication-procedure}
Given a received image $y$, LRW-Auth invokes the recovery module from Sec.~\ref{sec:image-recovery} (\frameworkfigref{fig:lrw-auth-framework}{LRW-Auth framework}). If recovery fails, authentication is rejected. Otherwise, the returned reference $x$ is compared with $y$:
\begin{equation}
    \mathrm{Auth}(y)=\mathrm{Compare}(y,x)\in\{0,1\}.
\label{eq:downstream-authentication}
\end{equation}
{\color{black}Here, $\mathrm{Compare}$ is a general term for a user-defined comparison procedure, such as a distance-based decision rule or a task-specific analyzer. It assesses the received image against the recovered original, returning $1$ for acceptance. The user determines which content to authenticate \textcolor{black}{and what transformations are acceptable}. These choices need not be fixed at generation time.}
\par}

\section{\textcolor{black}{Experiments}}
\label{sec:evaluation}

\textcolor{black}{We evaluate LRW through prompt compression, image quality, and message recovery, and examine its use in downstream authentication through LRW-Auth. Our instantiations demonstrate the framework with existing watermarking methods; measured quality and recovery rates reflect each method and its embedding configuration.}

\label{sec:experimental-setup}

\noindent\textbf{Datasets.}
We use MS-COCO 2017~\cite{lin2014coco} and Stable Diffusion Prompts (SDP)~\cite{gustavosta2022sdp} to evaluate generation quality, prompt compression, watermark robustness, and resistance to forgery attacks. W-Bench~\cite{lu2025wbench} provides source prompts and editing instructions for evaluating robustness to instruction-guided editing.

\smallskip
\noindent\textbf{Methods and implementation.}
We instantiate LRW with PRC~\cite{gunn2025undetectable}, Gaussian Shading~\cite{yang2024gaussian}, and T2SMark~\cite{yang2025t2smark}. We additionally evaluate SEAL~\cite{arabi2025seal} and SemBind's GS-S variant~\cite{zhang2026sembind} for image quality and forgery resistance. For watermark generation, all methods use Stable Diffusion 2.1 Base at $512\times512$, with 50 sampling steps and a guidance scale of 3. Watermark extraction uses 50-step DPM-Solver inversion based on the implementation of Hong et al.~\cite{hong2024exact}, with an empty prompt and a guidance scale of 3. Each method retains its watermark construction and semantic components. LRW uses Dream-v0-Base-7B~\cite{ye2025dream} with arithmetic coding~\cite{witten1987arithmetic} for lossless prompt compression and the first 64 bits of HMAC-SHA256 for message verification. \textcolor{black}{The seed and metadata use 8 and 20 bits, respectively.} {\color{black}Experiments are conducted on NVIDIA A40 GPUs.} \textcolor{black}{Implementation and settings: Appendices~\ref{app:algorithm-embedding} and~\ref{app:protocol}.}

{\color{black}
\paragraph{Threat model.}
\label{sec:threat-model}
The attacker has full access to the algorithms and generation model, along with watermarked images, but not the sender's authentication key $k_{\mathrm{ID}}$. It can generate and manipulate images locally, covering distortions, purification, reprompting, editing, and attempts to forge embedded recovery information. Verification is available through a remote interface that returns only acceptance or rejection, allowing the attacker to adapt subsequent attempts.
\par}

\subsection{Compressed Prompt Lengths}
\label{sec:compression-lengths}

\begin{figure}[!htbp]
\centering
\includegraphics[width=\columnwidth]{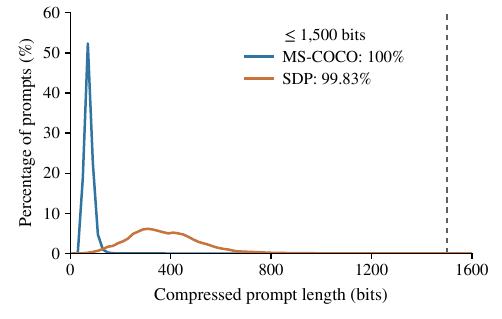}
\caption{\textbf{Distribution of losslessly compressed prompt lengths.} We evaluate \textcolor{black}{67,101} unique SDP prompts and \textcolor{black}{593,968} MS-COCO prompts. The 1,500-bit threshold accommodates 100\% of MS-COCO prompts and 99.83\% of SDP prompts.}
\label{fig:compression-lengths}
\end{figure}

\begin{figure*}[t]
\centering
\includegraphics[width=0.94\textwidth]{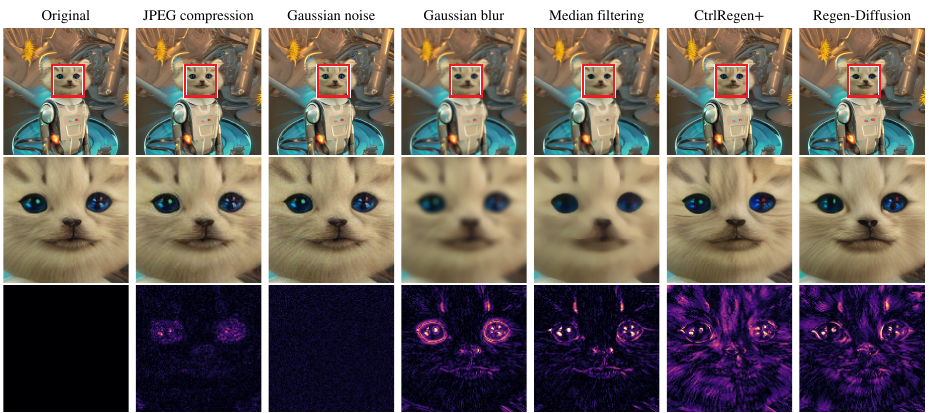}
\caption{\textbf{Visual effects of the evaluated distortions.} An LRW (T2SMark) image from SDP and its six transformed versions at the settings in Tables~\ref{tab:transmission-distortions} and~\ref{tab:purification}. Top: full images with the same region marked in red. Middle: fourfold enlargement of that region. Bottom: mean absolute RGB differences from the original region, shown on a shared linear color scale (8-bit intensity units, saturated at 64).}
\label{fig:distortion-effects}
\end{figure*}
We evaluate the space required to preserve prompts using the lossless compressor in Sec.~\ref{sec:wm-generation}. After exact text deduplication, we encode all \textcolor{black}{593,968} MS-COCO prompts and \textcolor{black}{67,101} SDP prompts. Figure~\ref{fig:compression-lengths} shows the resulting code-length distributions.

MS-COCO captions yield short, concentrated codes, with a mean of 73.09 bits and a standard deviation of 16.78 bits. SDP prompts produce longer and more variable codes, with a mean of 379.77 bits and a standard deviation of 163.42 bits. This difference reflects the concise scene descriptions in MS-COCO and the more detailed content and style specifications in SDP. Their median lengths are 71 and 361 bits, respectively, while their 99th percentiles are 125 and 840 bits. SDP has a sparse right tail extending to 3,259 bits. A 1,500-bit budget accommodates all MS-COCO prompts and 99.83\% of SDP prompts, with only \textcolor{black}{115} SDP prompts exceeding the threshold. Thus, lossless compression makes preserving the original prompt feasible within a compact watermark payload for the vast majority of inputs. \textcolor{black}{Distribution fits appear in Appendix~\ref{app:compression-results}.}

\subsection{Image Quality}
\label{sec:quality}
We assess the image quality of LRW instantiated with PRC, Gaussian Shading (GS), and T2SMark using FID~\cite{heusel2017fid} and CLIP score~\auditrev{\cite{radford2021clip,cherti2023reproducible}}. We compute FID using \textcolor{black}{5,000} images generated from MS-COCO prompts and the corresponding real images, and evaluate CLIP score on \textcolor{black}{1,000} SDP prompts and their generated images. Table~\ref{tab:instantiation-quality} shows that all three LRW instantiations maintain image quality comparable to unwatermarked generation \textcolor{black}{(Appendix~\ref{app:protocol-quality})}.

\begin{table}[t]
\centering
\caption{\textbf{Image quality of LRW instantiations.} Each LRW row applies the framework using the indicated watermarking method. All methods use the same generation settings.}
\label{tab:instantiation-quality}
\normalsize
\renewcommand{\arraystretch}{1.00}
\setlength{\tabcolsep}{2.5pt}
\begin{tabular*}{\columnwidth}{@{\extracolsep{\fill}}lrr@{}}
\toprule
Method & FID $\downarrow$ & CLIP score $\uparrow$ \\
\midrule
Unwatermarked & 20.00 & 0.3452 \\
\midrule
LRW (PRC~\cite{gunn2025undetectable}) & 20.49 & 0.3454 \\
LRW (GS~\cite{yang2024gaussian}) & 20.43 & 0.3464 \\
LRW (T2SMark~\cite{yang2025t2smark}) & 20.51 & 0.3467 \\
\midrule
{\color{black}SemBind~\cite{zhang2026sembind}} & 20.51 & 0.3461 \\
SEAL~\cite{arabi2025seal} & 20.66 & 0.3461 \\
\bottomrule
\end{tabular*}
\end{table}

\begin{table*}[t]
\centering
\caption{{\color{black}\textbf{LRW under transmission and processing distortions.} Message recovery success rate (\%). Success requires complete message recovery and valid MAC verification.}}
\label{tab:transmission-distortions}
\small
\renewcommand{\arraystretch}{1.08}
\setlength{\tabcolsep}{4pt}
\begin{tabular*}{\textwidth}{@{\extracolsep{\fill}}lrrrrrrrrrr@{}}
\toprule
& \multicolumn{5}{c}{MS-COCO} & \multicolumn{5}{c}{SDP} \\
\cmidrule(lr){2-6}\cmidrule(l){7-11}
Instantiation & Clean & \shortstack{JPEG\\compression} & \shortstack{Gaussian\\noise} & \shortstack{Gaussian\\blur} & \shortstack{Median\\filtering} & Clean & \shortstack{JPEG\\compression} & \shortstack{Gaussian\\noise} & \shortstack{Gaussian\\blur} & \shortstack{Median\\filtering} \\
\midrule
LRW (PRC~\cite{gunn2025undetectable}) & 100\% & 96\% & 99\% & 99\% & 93\% & 100\% & 96\% & 97\% & 99\% & 93\% \\
LRW (GS~\cite{yang2024gaussian}) & 100\% & 99\% & 99\% & 99\% & 95\% & 100\% & 93\% & 99\% & 74\% & 58\% \\
LRW (T2SMark~\cite{yang2025t2smark}) & 100\% & 100\% & 100\% & 100\% & 100\% & 100\% & 98\% & 99\% & 100\% & 93\% \\
\bottomrule
\end{tabular*}
\end{table*}

\subsection{Resistance to Message Forgery}
\label{sec:message-forgery}
The attacker retains a valid tag but modifies the compressed prompt, seed, or decoding metadata, then submits candidate images to the authentication service. The attack succeeds if the modified recovery information passes HMAC verification and causes LRW to reconstruct a different reference image.

We propose truncating HMAC-SHA256 to 64 bits to balance authentication overhead against watermark capacity. Let $q$ be the total number of HMAC verification attempts under the same key. Modeling the truncated HMAC as a secret random function, the probability of accepting recovery information that the sender has never authenticated is at most $q/2^{64}$~\cite{bellare2004verification,boneh2020graduate}. If the authentication service permits at most $2^{24}$ attempts under each key, this probability is at most $2^{-40}$. The limit applies to the combined attempts for all images and all verification servers using that key. Appendix~\ref{app:message-security} provides the security analysis.

\subsection{Robustness to Permissible Distortions}
\label{sec:permissible-distortions}
We define \emph{permissible distortions} as image transformations that preserve the content relevant to the application's authentication requirements. We evaluate common transmission and processing distortions and diffusion-based purification at settings intended to retain image content. Figure~\ref{fig:distortion-effects} illustrates the visual effects of all six transformations on the same watermarked image. Each condition uses \textcolor{black}{100} images from each dataset. For each LRW instantiation, success requires complete message recovery and valid MAC verification.

\paragraph{Transmission and processing distortions.}
We apply JPEG compression (quality 60), Gaussian noise ($\sigma=5/255$), Gaussian blur ($\sigma=3$, $7\times7$ kernel), and median filtering ($7\times7$ kernel). These operations model image changes introduced during transmission and processing. Table~\ref{tab:transmission-distortions} reports complete message recovery for the three LRW instantiations under these distortions.

\paragraph{Diffusion-based purification.}
Generative models can reconstruct a watermarked image after partial noising, seeking to suppress watermark signals while retaining image content~\cite{zhao2024removable,liu2025ctrlregen}. This differs from instruction-guided editing, which explicitly requests changes to image content. We evaluate CtrlRegen+~\cite{liu2025ctrlregen} at strength 0.1 and Regen-Diffusion~\cite{zhao2024removable} at noise timestep 60. Table~\ref{tab:purification} reports complete message recovery under both purification methods \textcolor{black}{(settings in Appendix~\ref{app:protocol-distortions})}.

\begin{table}[t]
\centering
\caption{{\color{black}\textbf{LRW under diffusion-based purification.} Message recovery success rate (\%), using the same criterion as Table~\ref{tab:transmission-distortions}. Ctrl: CtrlRegen+; Regen: Regen-Diffusion.}}
\label{tab:purification}
\small
\renewcommand{\arraystretch}{1.08}
\setlength{\tabcolsep}{2.5pt}
\begin{tabular*}{\columnwidth}{@{\extracolsep{\fill}}lrrrr@{}}
\toprule
& \multicolumn{2}{c}{MS-COCO} & \multicolumn{2}{c}{SDP} \\
\cmidrule(lr){2-3}\cmidrule(l){4-5}
Instantiation & Ctrl & Regen & Ctrl & Regen \\
\midrule
LRW (PRC~\cite{gunn2025undetectable}) & 95\% & 99\% & 95\% & 97\% \\
LRW (GS~\cite{yang2024gaussian}) & 96\% & 98\% & 63\% & 74\% \\
LRW (T2SMark~\cite{yang2025t2smark}) & 100\% & 100\% & 98\% & 99\% \\
\bottomrule
\end{tabular*}
\end{table}

\subsection{Forgery Attacks}
\label{sec:image-authentication}
\label{sec:forgery-attacks}
We evaluate reprompting~\cite{muller2025forgery}, which estimates the initial noise of a watermarked image $x$ and reuses it with a different prompt to generate new content $y$ carrying the watermark. We use the same \textcolor{black}{100} source--target pairs per dataset, with source images from MS-COCO and SDP and target prompts from I2P~\cite{schramowski2023safe}. The attack uses 50-step DDIM inversion \auditrev{followed by} 50-step generation with a guidance scale of 7.5 \textcolor{black}{(Appendix~\ref{app:protocol-forgery})}. {\color{black}We examine source-message recovery by LRW and authentication of the received image by LRW-Auth separately.}

{\color{black}
\paragraph{Message Recovery with LRW.}
Table~\ref{tab:lrw-reprompt-recovery} reports complete source-message recovery with valid MAC verification for each LRW instantiation. These rates measure whether the generation information needed to reconstruct the original image remains available after reprompting. The recovered original then provides reference evidence for LRW-Auth's content verification.
\par}

\begin{table}[t]
\color{black}
\centering
\caption{{\color{black}\textbf{LRW recovery under reprompting.} Complete source-message recovery with valid MAC verification (\%).}}
\label{tab:lrw-reprompt-recovery}
\small
\renewcommand{\arraystretch}{1.08}
\setlength{\tabcolsep}{2.5pt}
\begin{tabular*}{\columnwidth}{@{\extracolsep{\fill}}lrr@{}}
\toprule
Instantiation & MS-COCO & SDP \\
\midrule
LRW (PRC~\cite{gunn2025undetectable}) & \textcolor{black}{97\%} & \textcolor{black}{96\%} \\
LRW (GS~\cite{yang2024gaussian}) & \textcolor{black}{98\%} & \textcolor{black}{89\%} \\
LRW (T2SMark~\cite{yang2025t2smark}) & \textcolor{black}{99\%} & \textcolor{black}{99\%} \\
\bottomrule
\end{tabular*}
\end{table}

{\color{black}
\paragraph{Image Authentication with LRW-Auth.}
Table~\ref{tab:forgery-attacks} reports ROC-AUC ($\uparrow$) for distinguishing original watermarked images from reprompted forgeries using SemBind~\cite{zhang2026sembind} and SEAL~\cite{arabi2025seal}. \textcolor{black}{ROC curves: Appendix~\ref{app:forgery-roc}; examples: Appendix~\ref{app:qualitative-reprompting}.} SEAL bases forgery detection on caption-derived features, while SemBind uses learned visual semantic masks. Content changes can escape verification when these representations do not sufficiently distinguish the original and forged images. Moreover, a predefined semantic representation may omit details relevant to a downstream application's requirements.
\par}

\begin{table}[t]
\centering
\caption{\textbf{Forgery detection under reprompting.} \textcolor{black}{ROC-AUC ($\uparrow$) for SemBind and SEAL. LRW-Auth's forgery resistance follows from the two-case analysis below.}}
\label{tab:forgery-attacks}
\normalsize
\renewcommand{\arraystretch}{1.08}
\setlength{\tabcolsep}{2.5pt}
\begin{tabular*}{\columnwidth}{@{\extracolsep{\fill}}lrr@{}}
\toprule
Method & MS-COCO & SDP \\
\midrule
SemBind~\cite{zhang2026sembind} & 0.9962 & 0.9698 \\
SEAL~\cite{arabi2025seal} & 0.9772 & 0.9042 \\
\midrule
LRW-Auth & \multicolumn{2}{c}{\textcolor{black}{Forgery resistance by design}} \\
\bottomrule
\end{tabular*}
\end{table}

{\color{black}LRW-Auth guarantees that reprompting cannot bypass the application's content verification merely by retaining a valid watermark. \textcolor{black}{We analyze this guarantee through the two recovery outcomes below.} For reprompting that reuses the source message, (1) if extraction or MAC verification fails, LRW-Auth rejects $y$; (2) if the original message is recovered and verified, LRW reconstructs the original image $x$. With this complete reference image, users can assess the changes relevant to their application through $\mathrm{Compare}(y,x)$, using an appropriate analyzer or direct human inspection.} {\color{black}Once $\mathrm{Compare}(y,x)=0$ identifies a prohibited change, successful recovery also rejects the forgery. Together with rejection on recovery failure, this ensures that neither recovery outcome bypasses the user's content criterion.}

{\color{black}
\subsection{Editing Attacks}
\label{sec:editing-attacks}

\begin{figure}[t]
\centering
\includegraphics[width=\columnwidth]{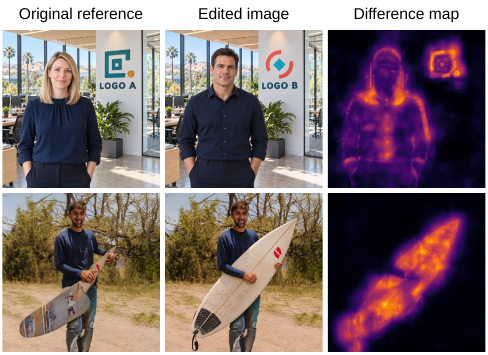}
\caption{{\color{black}\textbf{Reference-based comparison under image editing.} A joint identity and logo edit (top) and an object replacement (bottom) illustrate comparison against original references. Brighter regions indicate larger perceptual differences.}}
\label{fig:editing-authentication}
\end{figure}

Instruction-guided editing changes selected image content, allowing us to examine both LRW recovery and the application-dependent decisions of LRW-Auth. We apply InstructPix2Pix~\cite{brooks2023ip2p}, MagicBrush~\cite{zhang2023magicbrush}, and GPT Image using \textcolor{black}{100} source-prompt and editing-instruction pairs per editor from W-Bench~\cite{lu2025wbench}. Within each editor, these pairs are shared across all LRW instantiations \textcolor{black}{(settings in Appendix~\ref{app:protocol-editing})}.

\paragraph{Message Recovery with LRW.}
Table~\ref{tab:editing-recovery} reports complete source-message recovery with valid MAC verification. These results demonstrate LRW instantiated with existing watermarking methods and their selected configurations. LRW specifies the generation information required for original-image recovery, while the recovery rates reflect the underlying watermarking method, message length, and embedding configuration. \textcolor{black}{Qualitative comparisons appear in Appendix~\ref{app:qualitative-editing}.}

\begin{table}[t]
\color{black}
\centering
\caption{{\color{black}\textbf{LRW recovery under instruction-guided editing.} Complete source-message recovery with valid MAC verification (\%).}}
\label{tab:editing-recovery}
\small
\renewcommand{\arraystretch}{1.08}
\setlength{\tabcolsep}{2.5pt}
\begin{tabular*}{\columnwidth}{@{\extracolsep{\fill}}lrrr@{}}
\toprule
Instantiation & InstructPix2Pix & MagicBrush & GPT Image \\
\midrule
LRW (PRC~\cite{gunn2025undetectable}) & \textcolor{black}{86\%} & \textcolor{black}{72\%} & \textcolor{black}{64\%} \\
LRW (GS~\cite{yang2024gaussian}) & \textcolor{black}{86\%} & \textcolor{black}{83\%} & \textcolor{black}{70\%} \\
LRW (T2SMark~\cite{yang2025t2smark}) & \textcolor{black}{90\%} & \textcolor{black}{91\%} & \textcolor{black}{76\%} \\
\bottomrule
\end{tabular*}
\end{table}

\paragraph{Image Authentication with LRW-Auth.}
A change in semantic representation does not by itself determine whether an application should reject the edit. Changes irrelevant to the selected authentication objective may remain acceptable. LRW provides the complete original as reference evidence, allowing the verifier to choose which content to authenticate and which transformations to accept without fixing these choices at generation time. The recovered reference therefore supports diverse authentication tasks through task-specific analyzers or human inspection.

We visualize spatial LPIPS differences~\cite{zhang2018perceptual} between the original reference and the locally tampered image as heatmaps to indicate local perceptual changes.

\par}

\section{Conclusion}
\label{sec:conclusion}
We presented LRW, a lossless reversible watermarking framework that preserves compact generation information to recover the original watermarked image. Lossless prompt compression and deterministic reconstruction of the watermarked noise enable recovery without maintaining a per-image archive. HMAC protects the recovery information, while latent-space watermarking provides robustness to image transformations. Experiments across three watermarking methods characterize prompt compression, image quality, and message recovery under distortions, purification, reprompting, and editing. Using the recovered original as reference evidence, LRW-Auth supports downstream authentication requirements without predetermining which image properties to protect at generation time.

\label{main-text-end}
{\small
\bibliographystyle{ieeenat_fullname}
\bibliography{references}
}

\clearpage
\appendix
\makeatletter
\setlength{\@fptop}{0pt}
\setlength{\@dblfptop}{0pt}
\setlength{\@fpsep}{18pt}
\setlength{\@dblfpsep}{18pt}
\makeatother
\setcounter{page}{1}
\renewcommand{\thepage}{S\arabic{page}}
\setcounter{figure}{0}
\renewcommand{\thefigure}{S\arabic{figure}}
\setcounter{table}{0}
\renewcommand{\thetable}{S\arabic{table}}
\setcounter{equation}{0}
\renewcommand{\theequation}{S\arabic{equation}}
\renewcommand{\theHequation}{supp.\arabic{equation}}
\setcounter{algorithm}{0}
\renewcommand{\thealgorithm}{S\arabic{algorithm}}
\twocolumn[{
\centering
{\Large\bfseries Lossless Reversible Watermarking for AIGC\\
with Applications in Authentication\par}
\vspace{0.5em}
{\large\bfseries Supplementary Material\par}
\vspace{0.8em}
}]
{\color{black}
\noindent The supplementary material follows the order of the main paper. Section~\ref{app:algorithms} expands the method, and Section~\ref{app:protocol} gives the experimental settings. Sections~\ref{app:compression-results}--\ref{app:qualitative} provide details and additional results for each experiment in sequence.
\par}

\section{Method Details}
\raggedbottom

\label{app:algorithms}
\label{app:analysis}

{\color{black}
Algorithms~\ref{alg:lrw-embed}--\ref{alg:lrw-auth} detail Figs.~\ref{fig:framework} and~\ref{fig:lrw-auth-framework}. Here, $\mathrm{HMAC}$ includes the configured tag truncation.
\par}

\subsection{Watermark Construction and Embedding}
\label{app:algorithm-embedding}

\begin{algorithm}[H]
\color{black}
\caption{LRW: watermark construction and embedding}
\label{alg:lrw-embed}
\small
\begin{algorithmic}[1]
\Require Prompt $p$, seed, authentication key $k_{\mathrm{ID}}$
\Ensure Original watermarked image $x$
\State $(C,\mathrm{metadata})\gets\Compress(p)$
\State $\tau\gets\mathrm{HMAC}_{k_{\mathrm{ID}}}(\mathrm{metadata}\Vert\mathrm{seed}\Vert C)$
\State $M\gets\tau\Vert\mathrm{metadata}\Vert\mathrm{seed}\Vert C$
\State $r\gets G(\mathrm{seed})$
\State $s\gets\mathrm{WM}_{\mathrm{enc}}(M,r)$
\State $c\gets\mathrm{TextEnc}(p)$
\State $x\gets\mathrm{LDM}_{\Theta}(s,c)$
\State \Return $x$
\end{algorithmic}
\end{algorithm}

\paragraph{Lossless prompt compression.}
\label{app:implementation-compression}
{\color{black}
Dream-v0-Base-7B~\cite{ye2025dream} predicts token probabilities for arithmetic coding~\cite{witten1987arithmetic}. Tokens are encoded from left to right to produce the variable-length code $C$. Metadata stores the token count $N$ and code length $L=|C|$, specifying when text decoding stops and how many bits belong to $C$.

During decompression, the decoder uses the same LLM and coding settings to compute token probabilities from the tokens already recovered. Arithmetic decoding then recovers the next token from $C$. Repeating this process $N$ times recovers the original prompt.
\par}

\paragraph{Watermark instantiations.}
\label{app:implementation}
\label{app:implementation-carriers}
{\color{black}
We instantiate the watermark encoder and decoder with three methods. Separate positions in the initial noise carry the fixed-length and variable-length segments. This allows the decoder to extract the code length before decoding the variable-length segment.

\noindent\textit{Gaussian Shading.}
Gaussian Shading~\cite{yang2024gaussian} embeds bits by sampling within Gaussian intervals. Our implementation repeats the message bits, applies a shared private mask, and maps the resulting bits to Gaussian noise signs. Extraction reverses the mask and combines repeated observations by majority voting. The setup mask is fixed across images.

\noindent\textit{PRC.}
PRC~\cite{gunn2025undetectable} encodes the message with a pseudorandom error-correcting code and maps the coded bits to Gaussian noise signs. Extraction estimates these bits from inverted noise and applies error correction. We use a key-generation false-positive parameter of $10^{-9}$, posterior variance parameter 1.5, and at most 50 belief-propagation iterations.

\noindent\textit{T2SMark.}
T2SMark~\cite{yang2025t2smark} embeds bits through tail-truncated Gaussian sampling and keyed projection. The decoder reads the corresponding projections of the inverted noise. We use truncation parameter 0.674, a fixed 63-bit key for the fixed segment, and a 16-bit session value derived from the seed for the variable segment. This session value and all encoder randomness are regenerated during image recovery.
\par}

\subsection{Watermark Extraction and Image Recovery}
\label{app:algorithm-recovery}

{\color{black}
Algorithm~\ref{alg:lrw-recover} uses $\mathrm{WM}_{\mathrm{dec}}^{\mathrm{fix}}$ for the fixed-length fields and $\mathrm{WM}_{\mathrm{dec}}^{\mathrm{var}}$ for $C$. The fixed-length configuration is known in advance, whereas decoding $C$ requires the recovered length $\hat L$. For T2SMark, the recovered seed also determines the variable-length decoding configuration. The complete message must pass HMAC verification before image regeneration. Extraction, parsing, or decompression errors return \textsc{Fail}.
\par}

\begin{algorithm}[H]
\color{black}
\caption{LRW: watermark extraction and image recovery}
\label{alg:lrw-recover}
\small
\begin{algorithmic}[1]
\Require Received image $y$, authentication key $k_{\mathrm{ID}}$
\Ensure Reconstructed image $x$ on success; \textsc{Fail} otherwise
\State $\tilde s\gets\mathrm{Inv}_{\Theta}(y)$
\State $(\hat\tau,\widehat{\mathrm{metadata}},\widehat{\mathrm{seed}})\gets\mathrm{WM}_{\mathrm{dec}}^{\mathrm{fix}}(\tilde s)$
\State Read code length $\hat L$ from $\widehat{\mathrm{metadata}}$
\State $\hat C\gets\mathrm{WM}_{\mathrm{dec}}^{\mathrm{var}}(\tilde s;\hat L,\widehat{\mathrm{seed}})$
\State $\hat M\gets\hat\tau\Vert\widehat{\mathrm{metadata}}\Vert\widehat{\mathrm{seed}}\Vert\hat C$
\State $\tau'\gets\mathrm{HMAC}_{k_{\mathrm{ID}}}(\widehat{\mathrm{metadata}}\Vert\widehat{\mathrm{seed}}\Vert\hat C)$
\If{$\hat\tau\ne\tau'$}
    \State \Return \textsc{Fail}
\EndIf
\State $\hat p\gets\Decompress(\hat C,\widehat{\mathrm{metadata}})$
\If{decompression fails}
    \State \Return \textsc{Fail}
\EndIf
\State $\hat r\gets G(\widehat{\mathrm{seed}})$
\State $\hat s\gets\mathrm{WM}_{\mathrm{enc}}(\hat M,\hat r)$
\State $\hat c\gets\mathrm{TextEnc}(\hat p)$
\State $x\gets\mathrm{LDM}_{\Theta}(\hat s,\hat c)$
\State \Return $x$
\end{algorithmic}
\end{algorithm}

\paragraph{Exact image recovery.}
\label{app:exact-recovery}
{\color{black}
The estimated noise $\tilde s$ is used only for message extraction. Image generation instead uses $\hat s$, reconstructed by sampling from the recovered seed and embedding the verified message again. When $\hat M=M$, lossless decompression gives $\hat p=p$ and repeated watermark embedding gives $\hat s=s$. Consequently,
\begin{equation}
\hat M=M\quad\Longrightarrow\quad
\mathrm{LDM}_{\Theta}(\hat s,\hat c)
=\mathrm{LDM}_{\Theta}(s,c)=x.
\label{eq:supp-exact-recovery}
\end{equation}
This equality uses the same model weights, sampling settings, numerical execution, and image conversion. Exact diffusion inversion is unnecessary because the initial noise is reconstructed from the recovered message.
\par}

\subsection{LRW-Based Authentication}
\label{app:algorithm-authentication}

{\color{black}
Algorithm~\ref{alg:lrw-auth} uses LRW as the recovery module. The application \auditrev{can select} $\mathrm{Compare}$ after generation to specify the content and acceptance criterion, before evaluating candidate images.
\par}

\begin{algorithm}[H]
\color{black}
\caption{LRW-Auth: image authentication}
\label{alg:lrw-auth}
\small
\begin{algorithmic}[1]
\Require Received image $y$, authentication key $k_{\mathrm{ID}}$, application-defined $\mathrm{Compare}$
\Ensure Acceptance ($1$) or rejection ($0$)
\State $x\gets\mathrm{Recover}_{k_{\mathrm{ID}},\Theta}(y)$
\If{$x=\text{\textsc{Fail}}$}
    \State \Return $0$
\EndIf
\If{$\mathrm{Compare}(y,x)=1$}
    \State \Return $1$
\Else
    \State \Return $0$
\EndIf
\end{algorithmic}
\end{algorithm}

\begin{figure*}[t]
\centering
\includegraphics[width=0.94\textwidth]{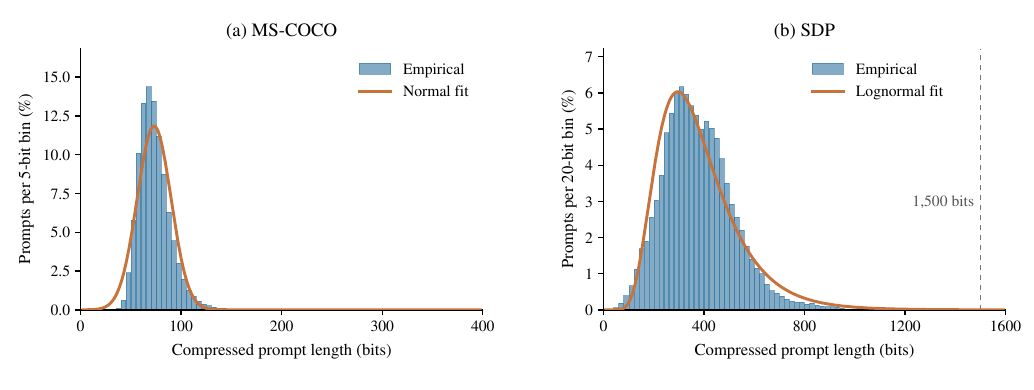}
\caption{\textcolor{black}{\textbf{Empirical and fitted compressed-prompt distributions.} Histograms of MS-COCO and SDP are shown with normal and lognormal fits, respectively. Bin widths are 5 and 20 bits. Fitting and normalization use all prompts; the SDP display ends at 1,600 bits. The dashed line marks 1,500 bits.}}
\label{fig:supp-compression}
\end{figure*}

\section{Experimental Settings}
\label{app:protocol}

\begin{table}[H]
\centering
\caption{Generation and extraction configuration.}
\label{tab:app-generation}
\small
\begin{tabular}{@{}ll@{}}
\toprule
Setting & Value \\
\midrule
Generation model & Stable Diffusion 2.1 Base \\
Image / latent dimensions & $512\times512$ / $4\times64\times64$ \\
Generation solver & DPM-Solver, order 1 \\
Generation steps / guidance & 50 / 3 \\
Inversion steps / guidance & 50 / 3 \\
Inversion prompt / order & Empty / 0 \\
Diffusion computation & FP32 \\
VAE latent refinement & 20 Adam updates \\
Learning rate / loss & 0.1 / summed squared error \\
Warmup / cosine horizon & 10 / 100 updates \\
Hardware & NVIDIA A40 GPU \\
\bottomrule
\end{tabular}
\end{table}

\begin{figure*}[t]
\centering
\includegraphics[width=0.92\textwidth]{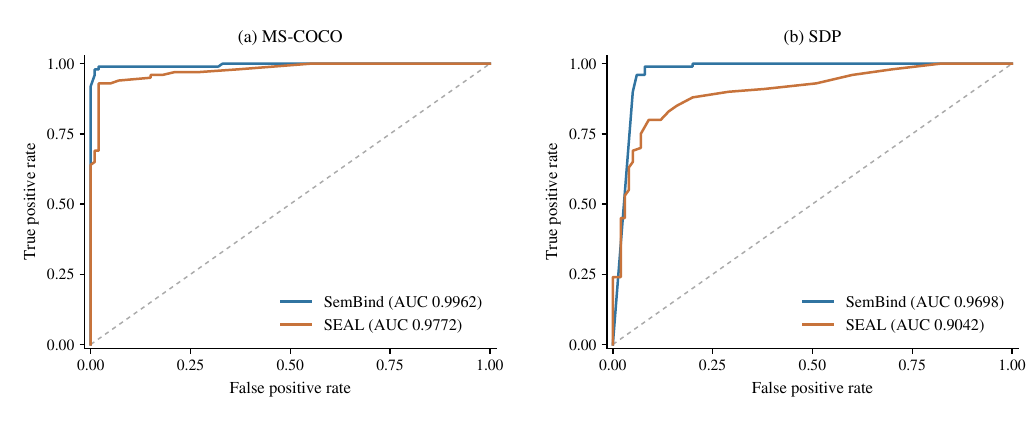}
\caption{\textbf{Empirical ROC curves under reprompting.} Original watermarked images are positives and reprompted images are negatives. Thus, the false positive rate is the fraction of reprompted images accepted at a score threshold. Curves use the saved per-image detection scores and reproduce the AUCs in Table~\ref{tab:forgery-attacks}.}
\label{fig:supp-forgery-roc}
\end{figure*}

\begin{figure*}[!t]
    \centering
    \includegraphics[width=0.94\textwidth]{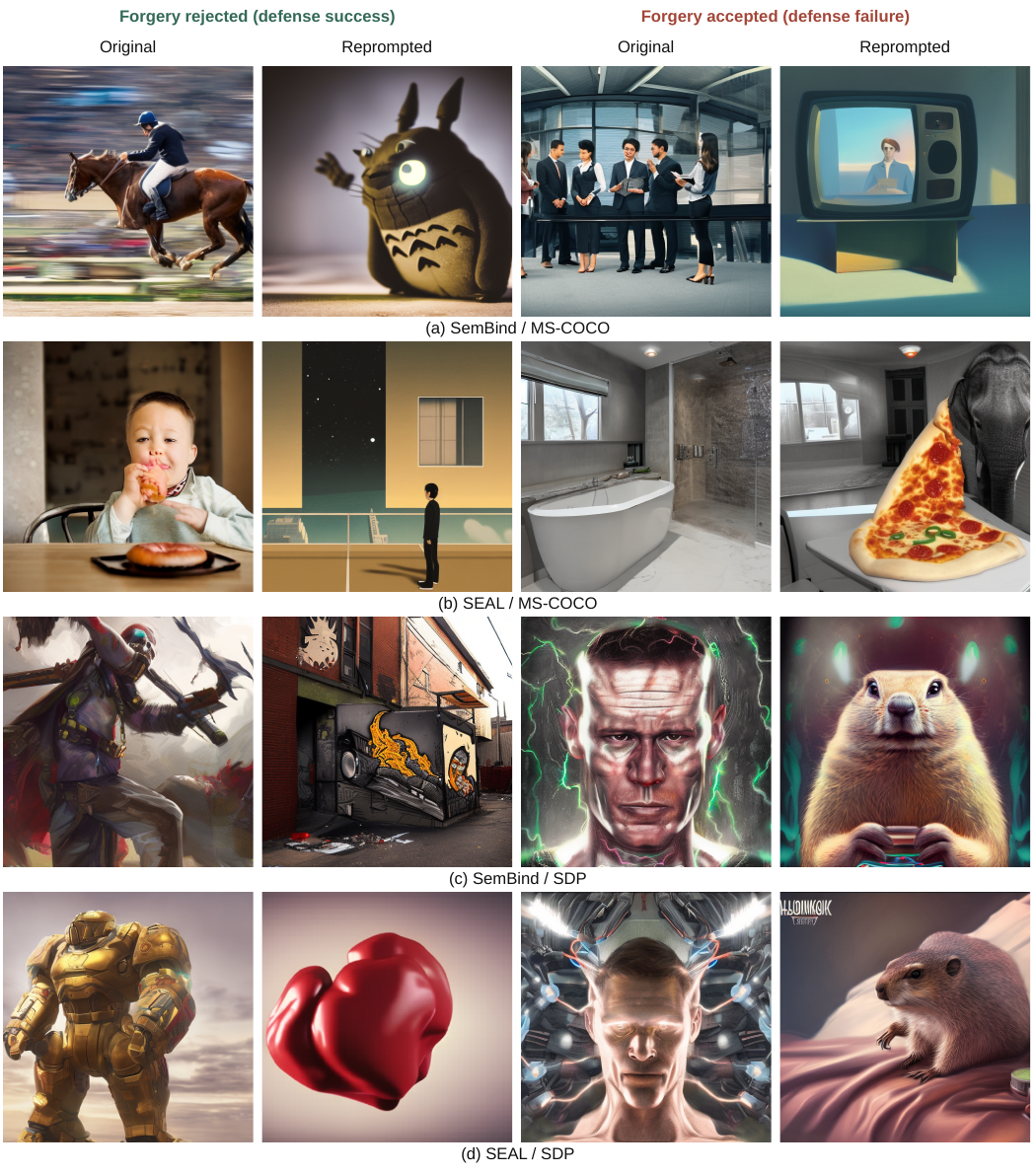}
    \caption{\textbf{Forgery detection by SemBind and SEAL.} \textcolor{black}{Rows (a)--(b) compare the two methods on MS-COCO, and rows (c)--(d) on SDP.} Each method is shown with one successfully rejected forgery (left) and one accepted forgery (right), alongside their original watermarked images. Detection thresholds are specified in Sec.~\ref{app:qualitative-reprompting}.}
    \label{fig:supp-reprompting}
\end{figure*}

\section{Compressed Prompt Lengths}
\label{app:additional-results}
\label{app:compression-results}
{\color{black}
Figure~\ref{fig:supp-compression} compares empirical compressed-prompt lengths with normal and lognormal fits for MS-COCO and SDP, respectively. Maximum-likelihood fitting gives $L_{\mathrm{COCO}}\sim\mathcal{N}(73.09,16.78^2)$ and $\ln L_{\mathrm{SDP}}\sim\mathcal{N}(5.858,0.411^2)$, where $L$ is the code length in bits. These fits predict a negligible probability of exceeding 1,500 bits for MS-COCO and approximately $0.020\%$ for SDP. The observed SDP distribution has a heavier tail, with $0.171\%$ of prompts exceeding this budget. Empirically, 1,500 bits cover all MS-COCO prompts and $99.83\%$ of SDP prompts, confirming coverage of nearly all evaluated prompts. Table~\ref{tab:supp-compression-statistics} reports the full statistics.
\par}

\begin{table}[H]
\centering
\caption{\textbf{Empirical compressed-prompt statistics.} All lengths are in bits; SD is the sample standard deviation.}
\label{tab:supp-compression-statistics}
\small
\setlength{\tabcolsep}{5pt}
\begin{tabular}{lrr}
\toprule
Statistic & MS-COCO & SDP \\
\midrule
Unique prompts & \textcolor{black}{593,968} & \textcolor{black}{67,101} \\
Mean & 73.09 & 379.77 \\
SD & 16.78 & 163.42 \\
Median & 71 & 361 \\
95th percentile & 102 & 628 \\
99th percentile & 125 & 840 \\
Maximum & 397 & 3,259 \\
Coverage at 1,500 bits & 100\% & 99.83\% \\
\bottomrule
\end{tabular}
\end{table}

\section{Image Quality}
\label{app:protocol-quality}
{\color{black}
FID~\cite{heusel2017fid} uses the 2,048-dimensional Inception features in \texttt{pytorch-fid}, with one caption per COCO validation image. CLIP score~\auditrev{\cite{radford2021clip,cherti2023reproducible}} is the mean normalized image--text cosine similarity from OpenCLIP ViT-g/14 with the LAION-2B checkpoint. Each method is evaluated once using the same prompts.

SemBind uses GS-S with the released 1,024-bit semantic hash model and mask ratio 1. SEAL uses its BLIP-2 captioner and caption embedding model, with 1,024 latent patches and seven-bit semantic codes. Both retain their original semantic construction and detection procedures.
\par}

\begin{figure*}[!t]
\centering
\includegraphics[width=0.97\textwidth]{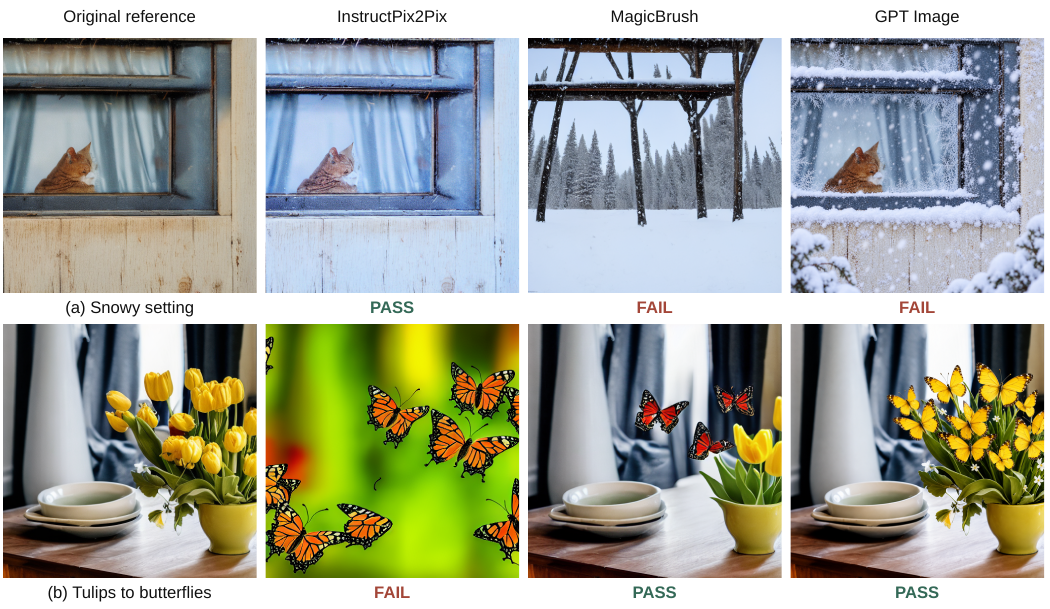}
\caption{\textbf{Editing a shared original with different editors.} Both rows use LRW (T2SMark). Within each row, all three editors receive the same original image and editing instruction. PASS and FAIL denote complete original-message recovery with a valid MAC. Every editor has one successful and one failed recovery.}
\label{fig:supp-editing}
\end{figure*}

\section{Resistance to Message Forgery}
\label{app:message-security}
The attacker modifies the compressed prompt, seed, or decoding metadata to make LRW reconstruct a different reference image. Without the sender's authentication key $k_{\mathrm{ID}}$, it tests candidate images through the remote authentication service. Message forgery succeeds if the service accepts recovery information that the sender has never authenticated.

Let $t$ be the tag length and $q$ the total number of HMAC verification attempts under the same key. Each candidate message and tag checked by the service counts as one attempt, including checks performed for different images or on different servers. In the ideal model, the truncated HMAC is a secret random function. The probability of a successful forgery within $q$ attempts is at most $q/2^t$, even when the attacker uses previous verification responses to choose subsequent attempts~\cite{bellare2004verification,boneh2020graduate}. Under the pseudorandom-function assumption for HMAC~\cite{bellare2006hmac}, the bound becomes
\[
\Pr[\text{message forgery}]
\leq \min\{1,q/2^t+\varepsilon_{\mathrm{HMAC}}\},
\]
where $\varepsilon_{\mathrm{HMAC}}$ is the attacker's advantage in distinguishing HMAC from a random function.

For the proposed 64-bit configuration, the service allows at most $2^{24}$ attempts in total under each key and rejects further requests under that key. The ideal forgery probability is then at most $2^{24}/2^{64}=2^{-40}$. Reusing recovery information already authenticated by the sender is not message forgery; image comparison determines whether its new carrier is acceptable.

\section{Robustness to Permissible Distortions}
\label{app:protocol-distortions}
{\color{black}We use JPEG compression at quality 60 with chroma subsampling, additive RGB Gaussian noise with standard deviation $5/255$, Gaussian blur with standard deviation 3 and a $7\times7$ kernel, and median filtering with a $7\times7$ kernel.\par}

CtrlRegen+~\cite{liu2025ctrlregen} uses Realistic Vision V4.0, the fine-tuned MSE VAE, UniPC sampling, guidance scale 2, and strength 0.1. This runs five denoising steps from a 50-step schedule. Canny thresholds are 100 and 150, both control scales are 1, and the output uses the implementation's color matching. Regen-Diffusion~\cite{zhao2024removable} uses SD2.1 Base, noise timestep 60, a 50-step schedule, guidance scale 7.5, and an empty prompt.

\section{Forgery Attacks}
\label{app:forgery}

\subsection{Reprompting Protocol}
\label{app:protocol-forgery}
Reprompting~\cite{muller2025forgery} uses direct VAE posterior-mean encoding and 50-step DDIM inversion with an empty prompt and guidance 1. The estimated noise and assigned target prompt then undergo 50-step DDIM sampling with guidance 7.5 and $\eta=0$. There is one candidate per source, without access to the source prompt, stored noise, recovery message, or private keys.

\subsection{Forgery Detection Curves}
\label{app:forgery-roc}
{\color{black}
Figure~\ref{fig:supp-forgery-roc} reports ROC curves using \textcolor{black}{100} original images and their \textcolor{black}{100} reprompted counterparts per dataset. SemBind uses recovered-bit accuracy after semantic unbinding~\cite{zhang2026sembind}; SEAL uses the number of semantic patches whose distance is below 2.3~\cite{arabi2025seal}. Higher scores favor original images. AUC is computed over all observed thresholds, with half credit for ties.
\par}

\subsection{Qualitative Examples}
\label{app:qualitative-reprompting}

{\color{black}
Figure~\ref{fig:supp-reprompting} shows successful and failed forgery rejection by SemBind and SEAL on both datasets. SemBind uses its native bit-accuracy threshold of $0.6484375$. SEAL uses a threshold of two matching semantic patches, calibrated on a separate set of unwatermarked images. A forged image accepted by the detector represents a failure to reject the forgery.
\par}

\section{Editing Attacks}
\label{app:qualitative}

\subsection{Editing Protocol}
\label{app:protocol-editing}
{\color{black}
InstructPix2Pix~\cite{brooks2023ip2p} and MagicBrush~\cite{zhang2023magicbrush} use 100 Euler-ancestral steps, text guidance 7.5, and image guidance 1.5. MagicBrush uses the released epoch-52 checkpoint. Editing instructions and random seeds are shared across LRW instantiations within each editor.

GPT editing uses \texttt{gpt-image-2.5-\allowbreak sunburst-\allowbreak 2026-09-08} with automatic quality and one opaque PNG per instruction. Outputs are converted to RGB and resized from $1024\times1024$ to $512\times512$ using LANCZOS. Each editor is evaluated on \textcolor{black}{100} source--instruction pairs.
\par}

\subsection{Qualitative Examples}
\label{app:qualitative-editing}
{\color{black}
Figure~\ref{fig:supp-editing} compares three editors using the same LRW (T2SMark) original and instruction within each row. Every editor is shown with one successful and one failed message recovery.
\par}

\clearpage

\end{document}